\documentclass[twocolumn,epjc3]{svjour3}
\RequirePackage{graphicx}
\RequirePackage{multirow}
\RequirePackage{yhmath}
\RequirePackage{hyperref}
\usepackage{color}
\usepackage{cases}
\usepackage{subfigure}
\usepackage{dcolumn,booktabs,bm}
\usepackage{slashed}
\usepackage{amsfonts,amssymb,stmaryrd,latexsym,amsmath}
\usepackage{textcomp}
\usepackage{widetext}

\journalname{Eur. Phys. J. C}

\begin{document}
\title{Magnetic moments of the spin-${3\over 2}$ doubly heavy baryons}
\author{Lu Meng\thanksref{e1,addr1},
        Hao-Song Li\thanksref{e2,addr1},
        Zhan-Wei Liu\thanksref{e3,addr2},
        Shi-Lin Zhu\thanksref{e4,addr1,addr3}}
\thankstext{e1}{e-mail: lmeng@pku.edu.cn}
\thankstext{e2}{e-mail: haosongli@pku.edu.cn}
\thankstext{e3}{e-mail: liuzhanwei@lzu.edu.cn}
\thankstext{e4}{e-mail: zhusl@pku.edu.cn}
\institute{School of Physics and State Key Laboratory of Nuclear
Physics and Technology, Peking University, Beijing 100871, China\label{addr1} \and School of Physical Science and Technology, Lanzhou
University, Lanzhou 730000, China \label{addr2}\and  Collaborative Innovation Center of Quantum Matter, Beijing 100871, China \label{addr3}}
\date{Received: date / Revised version: date}
%
\maketitle

\begin{abstract}
In this work, we investigate the chiral corrections to the magnetic moments of the
spin-$3\over 2$ doubly charmed baryons systematically up to next-to-next-to-leading order with the heavy
baryon chiral perturbation theory. The numerical results are given
up to next-to-leading order: $\mu_{\Xi^{*++}_{cc}}=1.72\mu_{N}$,
$\mu_{\Xi^{*+}_{cc}}=-0.09\mu_{N}$,
$\mu_{\Omega^{*+}_{cc}}=0.99\mu_{N}$. As a by-product, we have also
calculated the magnetic moments of the spin-$3\over 2$ doubly bottom
baryons and charmed bottom baryons:
 $\mu_{\Xi^{*0}_{bb}}=0.63\mu_{N}$,
$\mu_{\Xi^{*-}_{bb}}=-0.79\mu_{N}$,
$\mu_{\Omega^{*-}_{bb}}=0.12\mu_{N}$,
$\mu_{\Xi^{*+}_{bc}}=1.12\mu_{N}$,
$\mu_{\Xi^{*0}_{bc}}=-0.40\mu_{N}$,
$\mu_{\Omega^{*0}_{bc}}=0.56\mu_{N}$.
%
\end{abstract} 
\section{Introduction}\label{Sec1}

A doubly charmed baryon was first reported by the SELEX
Collaboration in the decay model
$\Xi^{+}_{cc}\rightarrow\Lambda_{c}^{+}K^{-}\pi^{+}$ with the mass
$M_{\Xi^{+}_{cc}}=3519\pm1\rm{MeV}$~\cite{Mattson:2002vu}. However,
no other collaborations confirmed the
observation~\cite{Ratti:2003ez,Aubert:2006qw,Chistov:2006zj}.
Recently, the spin-$1 \over 2$ doubly charmed baryon $\Xi_{cc}^{++}$
was reported by the LHCb Collaboration in the
$\Lambda_{c}^{+}K^{-}\pi^{+}\pi^{+}$ mass spectrum with the mass
$M_{\Xi^{++}_{cc}}$=3621.40$\pm$0.72 (stat.) $\pm$0.27(syst.)$\pm$0.14($\Lambda^{+}_{c}$)MeV~\cite{Aaij:2017ueg}.

The production of the doubly heavy baryons have been discussed in Refs.~\cite{Chang:2006eu,Chang:2006xp,Chang:2007xa,Chang:2007pp,Zheng:2015ixa,Chen:2014frw,Chen:2014hqa,Huan-Yu:2017emk}. Apart from the spin-$1 \over 2$ doubly charmed baryons, there exist
the spin-$3 \over 2$ doubly charmed baryons as degenerate states of
the spin-$1\over 2$ ones in the heavy quark limit. In the past
decades, the masses and other properties of spin-$1\over 2$ and
spin-$3\over 2$ doubly charmed baryons have been investigated
extensively in literature
\cite{Bagan:1992za,Roncaglia:1995az,SilvestreBrac:1996bg,Ebert:1996ec,Tong:1999qs,Itoh:2000um,Gershtein:2000nx,Kiselev:2001fw,Kiselev:2002iy,Narodetskii:2001bq,Lewis:2001iz,Ebert:2002ig,Mathur:2002ce,Flynn:2003vz,Vijande:2004at,Chiu:2005zc,Migura:2006ep,Albertus:2006ya,Martynenko:2007je,Tang:2011fv,Liu:2007fg,Roberts:2007ni,Patel:2007gx,Valcarce:2008dr,Liu:2009jc,Namekawa:2012mp,Alexandrou:2012xk,Aliev:2012ru,Namekawa:2013vu,Karliner:2014gca,Sun:2014aya,Chen:2015kpa,Sun:2016wzh,Shah:2016vmd,Chen:2016spr,Kiselev:2017eic,Chen:2017sbg,Hu:2005gf,Yu:2017zst,Li:2017ndo,Meng:2017udf,Guo:2017vcf,Wang:2017mqp,Wang:2017azm,Lu:2017meb,Xiao:2017udy}, where the mass splittings between the spin-$1 \over 2$ and
spin-$3\over 2$ doubly charmed baryons are from several tens MeV to
one hundred MeV. Since the real or virtual photons are usually used
as probes to explore the inner structures of baryons, the
electromagnetic form factors of the doubly charmed baryons are very
important. Especially the magnetic moments encode crucial
information of their underlying structure. In
Refs.~\cite{SilvestreBrac:1996bg,Patel:2007gx,Lichtenberg:1976fi,JuliaDiaz:2004vh,Faessler:2006ft,Dhir:2009ax,Branz:2010pq,Sharma:2010vv,Bose:1980vy,Bernotas:2012nz,Jena:1986xs,Oh:1991ws,Can:2013zpa,Li:2017cfz},
the magnetic moments of the spin-$1 \over 2$ and spin-$3 \over 2$
doubly charmed baryons have been investigated.

The magnetic moments of the spin-$3\over 2$ doubly charmed baryons were
first investigated in the naive quark model by
Lichtenberg~\cite{Lichtenberg:1976fi}. After that, many authors have
employed the quark model to calculate the spin-$3\over 2$ doubly
charmed baryon magnetic
moments~\cite{Albertus:2006ya,Patel:2007gx,Sharma:2010vv}. Apart
from the quark models, the magnetic moments of the spin-$3 \over 2$
doubly charmed baryons have been predicted employing the effective
quark mass and screened charge scheme in Ref.~\cite{Dhir:2009ax}.
The magnetic moments of the spin-$3 \over 2$ doubly charmed baryons have
been also calculated in the MIT bag
model~\cite{Bose:1980vy,Bernotas:2012nz} and Skymion
model~\cite{Oh:1991ws}.

Compared with the quark model, the chiral perturbation theory
(ChPT)~\cite{Weinberg:1978kz,Scherer:2002tk} provides a systematic
framework to calculate the electromagnetic form factors of the
baryons order by order. However, the baryon mass does not vanish in the
chiral limit, which introduces another energy scale besides the
$\Lambda_{\chi}=4\pi f_\pi$. In order to employ ChPT in the baryon
sector, the heavy baryon chiral perturbation theory (HBChPT) was
proposed~\cite{Jenkins:1990jv,Jenkins:1992pi,Bernard:1992qa,Bernard:1995dp}.
The magnetic moments of the octet, decuplet and spin-$1 \over
2$ doubly charmed baryons have been calculated in
the framework of HBChPT~\cite{Jenkins:1992pi,Meissner:1997hn,Li:2016ezv,Li:2017cfz}. The
decuplet to
octet and spin-$3 \over 2$ to
spin-$1 \over 2$ doubly charmed baryons transition magnetic moments were also
investigated~\cite{Jenkins:1990jv,Hemmert:1996xg,Gellas:1998wx,Gail:2005gz,Li:2017vmq,Li:2017pxa}.

In this paper, we investigate the magnetic moments of the spin-$3\over 2$ doubly heavy
baryons within the framework of HBChPT. We use the quark
model to estimate the low energy constants (LECs), since there does
not exist any experiment data. The numerical results are given to the
next-to-leading order while the analytical results are presented to
the next-to-next-to-leading order.

We first discuss the electromagnetic form factors of the
spin-$3\over 2$ baryons in Sec.~\ref{SecEM}. The chiral
Lagrangians are constructed in Sec.~\ref{SecLag}. In
Sec.~\ref{SecAnaly}, we calculate the magnetic moments analytically
order by order. In Sec.~\ref{SecNO}, with the help of quark model,
we estimate the LECs and give the numerical results of the magnetic
moments to the next-to-leading order. A short summary is given in
Sec.~\ref{SecConcld}. All the coefficients of the loop corrections
are collected in the \ref{AppCG}.

\section{Electromagnetic form factors of the spin-$\frac{3}{2}$ doubly charmed baryons }\label{SecEM}

For the spin-$\frac{3}{2}$ doubly charmed baryons, one can
parameterize the general electromagnetic current matrix
elements~\cite{Nozawa:1990gt}, which satisfy the gauge invariance,
parity conservation and time-reversal invariance:
\begin{equation}
\langle
T(p^{\prime})|J_{\mu}|T(p)\rangle=\bar{u}^{\rho}(p^{\prime})O_{\rho\mu\sigma}(p^{\prime},p)u^{\sigma}(p),
\end{equation}
with
\begin{equation}
\begin{split}
O_{\rho\mu\sigma}(p^{\prime},p)=&-g_{\rho\sigma}(A_{1}\gamma_{\mu}+\frac{A_{2}}{2M_{T}}P_{\mu})\\
&-\frac{q_{\rho}q_{\sigma}}{(2M_{T})^{2}}(C_{1}\gamma_{\mu}+\frac{C_{2}}{2M_{T}}P_{\mu}),
\end{split}
\end{equation}
where $p$ and $p'$ are the momenta of the spin-$3\over 2$ doubly
charmed baryons. $P=p+p'$, $q=p'-p$. $M_T$ is the doubly charmed
baryon mass and $u_{\sigma}$ is the Rarita-Schwinger
spinor~\cite{Rarita:1941mf}.

In the heavy baryon limit, the baryon field can be decomposed into
the large component $\mathcal{T}$ and small component $\mathcal{H}$,
$$T=e^{-iM_Tv\cdot x}(\mathcal{T}+\mathcal{H})$$
$$\mathcal{T}=e^{iM_Tv\cdot x}{1+\slashed{v}\over 2}T$$
$$\mathcal{H}=e^{iM_Tv\cdot x}{1-\slashed{v}\over 2}T$$
where $v_\mu$ is the velocity of the baryon. In the heavy baryon
limit, the matrix elements of the electromagnetic current $J_{\mu}$
can be re-parametrized as \cite{Li:2016ezv}
 \begin{equation}
\langle
\mathcal{T}(p^{\prime})|J_{\mu}|\mathcal{T}(p)\rangle=\bar{u}^{\rho}(p^{\prime})\mathcal{O}_{\rho\mu\sigma}(p^{\prime},p)u^{\sigma}(p),
\end{equation}
with
 \begin{equation}
 \begin{split}
 \mathcal{O}_{\rho\mu\sigma}(p^{\prime},p)&=-g_{\rho\sigma}\left[v_{\mu}F_{1}(q^{2})+\frac{[S_{\mu},S_{\alpha}]}{M_{T}}q^{\alpha}F_{2}(q^{2})\right]\\
 &-\frac{q^{\rho}q^{\sigma}}{(2M_{T})^{2}}\left[v_{\mu}F_{3}(q^{2})+\frac{[S_{\mu},S_{\alpha}]}{M_{T}}q^{\alpha}F_{4}(q^{2})\right]
 \end{split}
 \end{equation}
where $S_{\mu}={i\over 2}\gamma_5\sigma_{\mu\nu}v^\nu$ is the
covariant spin-operator. The charge (E0), electro-quadrupole (E2),
magnetic-dipole (M1), and magnetic octupole (M3) form factors read
 \begin{eqnarray}
 G_{E0}(q^{2})&=&(1+\frac{2}{3}\tau)[F_{1}+\tau(F_{1}-F_{2})]\nonumber\\
 &&-\frac{1}{3}\tau(1+\tau)[F_{3}+\tau(F_{3}-F_{4})],\\
 G_{E2}(q^{2})&=&[F_{1}+\tau(F_{1}-F_{2})]\nonumber \\
 &&-\frac{1}{2}(1+\tau)[F_{3}+\tau(F_{3}-F_{4})],\\
G_{M1}(q^{2})&=&(1+\frac{4}{5}\tau)F_{2}-\frac{2}{5}\tau(1+\tau)F_{4},\\
 G_{M3}(q^{2})&=&F_{2}-\frac{1}{2}(1+\tau)F_{4}.
  \end{eqnarray}
where $\tau=-\frac{q^{2}}{(2M_{T})^{2}}$. When $q^2=0$, we obtain
the magnetic moment $\mu_{T}=G_{M1}(0){e\over 2M_T}$.

\section{Chiral Lagrangians}\label{SecLag}

\subsection{The leading order chiral Lagrangians}
To calculate the chiral corrections to the magnetic moments, we
construct the relevant chiral Lagrangians. The doubly charmed baryon
fields read
\begin{equation}
B=\left(\begin{array}{c}
\Xi_{cc}^{++}\\
\Xi_{cc}^{+}\\
\Omega_{cc}^{+}
\end{array}\right), T^{\mu}=\left(\begin{array}{c}
\Xi_{cc}^{*++}\\
\Xi_{cc}^{*+}\\
\Omega_{cc}^{*+}
\end{array}\right)^\mu,\Rightarrow \left(\begin{array}{c}
ccu\\
ccd\\
ccs
\end{array}\right).
\end{equation}
where the $B$ and $T^{\mu}$ are spin-${1\over 2}$ and spin-${3\over
2}$ doubly chamed baryon fields respectively. We follow the
notations in Refs.~\cite{Li:2016ezv,Scherer:2002tk,Bernard:1995dp}
to define the basic chiral effective Lagrangians of the pseudoscalar
mesons. The pseudoscalar meson fields are introduced as follows,
\begin{equation}
\phi=\left(\begin{array}{ccc}
\pi^{0}+\frac{1}{\sqrt{3}}\eta & \sqrt{2}\pi^{+} & \sqrt{2}K^{+}\\
\sqrt{2}\pi^{-} & -\pi^{0}+\frac{1}{\sqrt{3}}\eta & \sqrt{2}K^{0}\\
\sqrt{2}K^{-} & \sqrt{2}\bar{K}^{0} & -\frac{2}{\sqrt{3}}\eta
\end{array}\right).
\end{equation}
 The chiral connection and axial vector
field are defined as~\cite{Scherer:2002tk,Bernard:1995dp},
\begin{equation}
\Gamma_{\mu}=\frac{1}{2}\left[u^{\dagger}(\partial_{\mu}-ir_{\mu})u+u(\partial_{\mu}-il_{\mu})u^{\dagger}\right],
\end{equation}
\begin{equation}
u_{\mu}=
i\left[u^{\dagger}(\partial_{\mu}-ir_{\mu})u-u(\partial_{\mu}-il_{\mu})u^{\dagger}\right],
\end{equation}
where
\begin{eqnarray}
&&u^{2}=\mathit{U}=\exp(i\phi/f_{0})\\
&&r_{\mu}=l_{\mu}=-eQA_{\mu},
\end{eqnarray}
For the Lagrangians with the baryon fields, $Q=Q_B=\rm{diag}(2,1,1)$ and
for the pure meson Lagrangians $Q=Q_M=\rm{diag}(2/3,-1/3,-1/3)$.
$f_0$ is the decay constant of the pseudoscalar meson in the chiral
limit. The experimental value of the pion decay constant
$f_{\pi}\approx$ 92.4 MeV while $f_{K}\approx$ 113 MeV,
$f_{\eta}\approx$ 116 MeV.

The leading order ($\mathcal{O}(p^{2})$) pure meson Lagrangian is
\begin{eqnarray}
&&\mathcal{L}_{\pi\pi}^{(2)}=\frac{f_{0}^{2}}{4}{\rm
Tr}[\nabla_{\mu}U(\nabla^{\mu}U)^{\dagger}] \label{Lag:meson1},\\
&&\nabla_{\mu}U=\partial_{\mu}U-ir_{\mu}U+iUl_{\mu},
\end{eqnarray}
where the superscript denotes the chiral order.
The leading order doubly charmed baryon Lagrangians and meson-baryon
interaction Lagrangians read
\begin{eqnarray}
{\cal L}_{TT}^{(1)}&=&\bar{T}^{\mu}[-g_{\mu\nu}(i\slashed{D}-M_{T})+i(\gamma_{\mu}D_{\nu}+\gamma_{\nu}D_{\mu})\nonumber  \\
&-&\gamma_{\mu}(i\slashed{D}+M_{T})\gamma_{\nu}]T^{\nu} 
+\frac{H}{2}\left(\bar{T}^{\mu}g_{\mu\nu}\slashed{u}\gamma_{5}T^{\nu}\right), \label{Lag:op1tt} \\
\mathcal{L}_{BB}^{(1)}&=&\bar{B}(i\slashed{D}-M_{B}+\frac{\tilde{g}_{A}}{2}\gamma^{\mu}\gamma_{5}u_{\mu})B ,\label{Lag:op1bb}\\
\mathcal
{L}_{BT}^{(1)}&=&\frac{C}{2}\left(\bar{T}^{\mu}u_{\mu}B+\bar{B}u_{\mu}T^{\mu}\right).
\label{Lag:op1bt}
\end{eqnarray}
We use the subscript to denote the two particles involved in the
Lagrangians. $M_B$ is the spin-${1\over 2 }$ doubly charmed baryon
mass. $\tilde{g}_A$, $C$ and $H$ are the coupling constants. The
covariant derivative is defined as $D_{\mu}\equiv
\partial_{\mu}+\Gamma_{\mu}$. Both $\mathcal{L}_{TT}$ and
$\mathcal{L}_{BB}$ contain the free and interaction terms.

In the framework of HBChPT, the leading order nonrelativistic
Lagrangians read
\begin{eqnarray}
{\cal L}_{TT}^{(1)}&=&\bar{{\cal T}}^{\mu}\left[-iv\cdot Dg_{\mu\nu}\right]{\cal T}^{\nu}+H\left(\bar{{\cal T}}^{\mu}g_{\mu\nu}u\cdot S{\cal T}^{\nu}\right), \label{Lag:op1rdtt} \\
{\cal L}_{BT}^{(1)}&=&\frac{C}{2}\left(\bar{{\cal T}}^{\mu}u_{\mu}{\cal B}+\bar{{\cal B}}u_{\mu}{\cal T}^{\mu}\right),\label{Lag:op1rdbt} \\
{\cal L}_{BB}^{(1)}&=&{\cal \bar{B}}i(v\cdot D-\delta){\cal B}+{\cal
\bar{B}}\tilde{g}_{A}S\cdot u{\cal B}, \label{Lag:op1rdbb}
\end{eqnarray}
where $\delta\equiv M_B-M_T$. Since the spin-$3\over 2$ doubly
charmed baryons have not been observed in the experiments, we take two values the mass splitting,
$\delta=$-100 MeV and $\delta=$-70 MeV, in our work. We do not consider the mass
difference among the doubly charmed baryon triplets. The coupling
constants are estimated with the help of quark model in
Refs.~\cite{Li:2017cfz,Li:2017pxa}, $\tilde{g}_{A}=-{1\over
5}g_N=-0.25$, $C=-\frac{2\sqrt{3}}{5}g_{N}=-0.88 $ and
$H=-\frac{3}{5}g_{N}=-0.76 $, where $g_N=1.267$ is the nucleon axial
charge. For the pseudoscalar meson masses, we use $m_{\pi}=0.140$
GeV, $m_{K}=0.494$ GeV, and $m_{\eta}=0.550$ GeV. We use the nucleon
mass $M_N=0.938\rm{GeV}$.

\subsection{The next-to-leading order chiral Lagrangians}

The $\mathcal{O}(p^{2})$ Lagrangians contribute to the magnetic
moments of the spin-${3\over 2}$ doubly charmed baryons
\begin{equation}
\mathcal{L}_{TT}^{(2)}=\frac{-ib_{1}^{tt}}{2M_{T}}\bar{T}^{\mu}\hat{F}_{\mu\nu}^{+}T^{\nu}+\frac{-ib_{2}^{tt}}{2M_{T}}\bar{T}^{\mu}\langle
F_{\mu\nu}^{+}\mathcal{\rangle}T^{\nu},  \label{Lag:op2tt}
\end{equation}
\begin{equation}
{\cal
L}_{BB}^{(2)}=\frac{b_{1}^{bb}}{8M_{T}}\bar{B}\sigma^{\mu\nu}\hat{F}_{\mu\nu}^{+}B+\frac{b_{2}^{bb}}{8M_{T}}\bar{B}\sigma^{\mu\nu}\langle
F_{\mu\nu}^{+}\rangle B,  \label{Lag:op2bb}
\end{equation}
\begin{eqnarray}
&&{\cal L}_{BT}^{(2)}=b_{1}^{bt}\frac{i}{4M_{T}}\bar{B}\hat{F}_{\mu\nu}^{+}\gamma^{\nu}\gamma_{5}T^{\mu}+b_{2}^{bt}\frac{i}{4M_{T}}\bar{B}\langle F_{\mu\nu}^{+}\rangle\gamma^{\nu}\gamma_{5}T^{\mu} \nonumber \\
&&-b_{1}^{bt}\frac{i}{4M_{T}}\bar{T}^{\mu}\hat{F}_{\mu\nu}^{+}\gamma^{\nu}\gamma_{5}B-b_{2}^{bt}\frac{i}{4M_{T}}\bar{T^{\mu}}\langle
F_{\mu\nu}^{+}\rangle\gamma^{\nu}\gamma_{5}B, \label{Lag:op2bt}
\end{eqnarray}
where the coefficients $b_{1,2}^{tt,bb,bt}$ are the LECs which
contribute to the magnetic moments at tree level. The chiral
covariant QED field strength tensors $F_{\mu\nu}^{\pm}$ are defined as
\begin{eqnarray}
F_{\mu\nu}^{\pm} & = & u^{\dagger}F_{\mu\nu}^{R}u\pm
uF_{\mu\nu}^{L}u^{\dagger},\\
F_{\mu\nu}^{R} & = &
\partial_{\mu}r_{\nu}-\partial_{\nu}r_{\mu}-i[r_{\mu},r_{\nu}],\\
F_{\mu\nu}^{L} & = &
\partial_{\mu}l_{\nu}-\partial_{\nu}l_{\mu}-i[l_{\mu},l_{\nu}].
\end{eqnarray}
Since the $Q_B$ is not traceless, the operator ${F}_{\mu\nu}^{+}$
can be divided into two parts, $\hat{F}_{\mu\nu}^{+}$ and
$\langle{F}_{\mu\nu}^{+}\rangle$. $\langle F^+_{\mu\nu}\rangle
\equiv \text{Tr}(F^+_{\mu\nu})$. The operator
$\hat{F}_{\mu\nu}^{+}=F_{\mu\nu}^{+}-\frac{1}{3}\rm \langle
F_{\mu\nu}^{+}\rangle$ is traceless and transforms as the adjoint
representation under the chiral transformation. Recall that the
direct product of the representation of SU(3) group $3\otimes\bar{3}
= 1\oplus8$. Therefore, there are two independent interaction terms
in the $\mathcal{O}(p^{2})$ Lagrangians for the magnetic moments of
the doubly charmed baryons. The nonrelativistic Lagrangians
corresponding to Eqs.~(\ref{Lag:op2tt}-\ref{Lag:op2bt}) are:
\begin{eqnarray}
&\mathcal{L}_{TT}^{(2)}=\frac{-ib_{1}^{tt}}{2M_{T}}\bar{\mathcal{T}}^{\mu}\hat{F}_{\mu\nu}^{+}\mathcal{T}^{\nu}+\frac{-ib_{2}^{tt}}{2M_{T}}\bar{\mathcal{T}}^{\mu}\langle F_{\mu\nu}^{+}\mathcal{\rangle T}^{\nu}, \label{Lag:op2rdtt} \\
&{\cal L}_{BT}^{(2)}=b_{1}^{bt}\frac{i}{2M_{T}}\bar{{\cal B}}\hat{F}_{\mu\nu}^{+}S^{\nu}{\cal T}^{\mu}+b_{2}^{bt}\frac{i}{2M_{T}}\bar{{\cal B}}\langle F_{\mu\nu}^{+}\rangle S^{\nu}{\cal T}^{\mu} \nonumber \\
&-b_{1}^{bt}\frac{i}{2M_{T}}\bar{{\cal T}}^{\mu}\hat{F}_{\mu\nu}^{+}S^{\nu}{\cal B}-b_{2}^{bt}\frac{i}{2M_{T}}\bar{{\cal T}^{\mu}}\langle F_{\mu\nu}^{+}\rangle S^{\nu}{\cal B}, \label{Lag:op2rdbt}\\
&{\cal L}_{BB}^{(2)}=-\frac{ib_{1}^{bb}}{4M_{T}}\bar{{\cal
B}}[S^{\mu},S^{\nu}]\hat{F}_{\mu\nu}^{+}{\cal B}-{\cal
\bar{B}}\frac{ib_{2}^{bb}}{4M_{T}}[S^{\mu},S^{\nu}]\langle
F_{\mu\nu}^{+}\rangle{\cal B}, \nonumber \\\label{Lag:op2rdbb}
\end{eqnarray}

We also need the second order pseudoscalar meson and doubly charmed
baryon interaction Lagrangians
\begin{eqnarray} {\cal
L}_{TT}^{(2)}=&&\frac{ig_{1}^{tt}}{4M_{T}}\bar{T}^{\mu}\{u_{\rho},u_{\sigma}\}\sigma^{\rho\sigma}g_{\mu\nu}T^{\nu}\nonumber\\
&&+\frac{ig_{2}^{tt}}{4M_{T}}\bar{T}^{\mu}[u_{\rho},u_{\sigma}]\sigma^{\rho\sigma}g_{\mu\nu}T^{\nu},\label{Lag:op2g}
\end{eqnarray}
where $g_{1,2}^{tt}$ are the coupling constants. Recall that for
SU(3) group representations,
\begin{eqnarray}
3\otimes\bar{3} & = & 1\oplus8\label{Eq:flavor1},\\
8\otimes8 & = &
1\oplus8_{1}\oplus8_{2}\oplus10\oplus\bar{10}\oplus27.\label{Eq:flavor2}
\end{eqnarray}
Both $u_{\mu}$ and $u_{\nu}$ transform as the adjoint
representation. The two terms in Eq.~(\ref{Lag:op2g}) correspond to
the product of $u_{\mu}$ and $u_{\nu}$ belonging to the $8_1$ and
$8_2$ flavor representations, respectively. The $g_{1}^{tt}$ term
vanishes because of the anti-symmetric Lorentz structure. Thus,
there is only one linearly independent LEC $g_{2}^{tt}$, which
contributes to the spin-$3\over 2$ doubly charmed baryon magnetic
moments up to $\mathcal{O}(p^3)$. The second order pseudoscalar
meson and baryon nonrelativistic Lagrangian reads
\begin{eqnarray}
{\cal L}_{TT}^{(2)}=\frac{g_{2}^{tt}}{2M_{T}}\bar{{\cal
T}}^{\mu}[S^{\rho},S^{\sigma}][u_{\rho},u_{\sigma}]g_{\mu\nu}{\cal
T}^{\nu} \label{Lag:op2grd}
\end{eqnarray}
The above Lagrangian contributes to the doubly charmed baryon
magnetic moments in the diagram (j) of the Fig.~\ref{fig:allloop}.

\subsection{The higher order chiral Lagrangians }

To calculate the $\mathcal{O}(p^{3})$ magnetic moments at the tree
level, we also need the $\mathcal{O}(p^{4})$ electromagnetic chiral
Lagrangians. The possible flavor structures are listed in Table
\ref{Table:Flavor structure}, where $\chi^{+}$=diag(0,0,1) at the
leading order. Recalling the flavor representation in
Eqs.~(\ref{Eq:flavor1}), (\ref{Eq:flavor2}), the leading order term
of the operator $[\hat{F}_{\mu\nu}^{+},\hat{\chi}_{+}]$ vanishes
after expansion since both $F^{+}_{\mu\nu}$ and $\chi^+$ are
diagonal. Meanwhile, the $\langle
F_{\mu\nu}^{+}\rangle\langle\chi_{+}\rangle$ and $\langle
F_{\mu\nu}^{+}\rangle\hat{\chi}_{+}$ terms can be absorbed into
Eq.~(\ref{Lag:op2tt}) by renormalizing  the LECs $b_1^{tt}$ and
$b_2^{tt}$. Thus, the independent $\mathcal{O}(p^4)$ Lagrangians
read:
\begin{equation}
\begin{split}
{\cal {\cal L}}^{(4)}=&\frac{-ia_{1}}{2M_{T}}\bar{T}^{\mu}\langle
F_{\mu\nu}^{+}\rangle\hat{\chi}_{+}T^{\nu}
+\frac{-ia_{2}}{2M_{T}}\bar{T}^{\mu}\langle\hat{F}_{\mu\nu}^{+}\hat{\chi}_{+}\rangle
T^{\nu}\\
&+\frac{-ia_{3}}{2M_{T}}\bar{T}^{\mu}\{\hat{F}_{\mu\nu}^{+},\hat{\chi}_{+}\}\hat{\chi}_{+}T^{\nu}.
\label{op4r}
\end{split}
\end{equation}

\begin{table*}
\caption{The possible flavor structures of the $\mathcal{O}(p^4)$
Lagrangians which contribute to the magnetic moments. }
\label{Table:Flavor structure}
  \centering
\begin{tabular}{c|c|c|c|c|c|c}
\hline\noalign{\smallskip} Group representation &
$1\otimes1\rightarrow1$ & $1\otimes8\rightarrow8$ &
$8\otimes1\rightarrow8$ & $8\otimes8\rightarrow1$ &
$8\otimes8\rightarrow8_{1}$ &
$8\otimes8\rightarrow8_{2}$\\
 \noalign{\smallskip}\hline\noalign{\smallskip}
  Flavor structure 
&$\langle F_{\mu\nu}^{+}\rangle \langle\chi_{+}\rangle$ & $\langle
F_{\mu\nu}^{+}\rangle \hat{\chi}_{+}$ &
$\hat{F}_{\mu\nu}^{+}\langle\chi_{+}\rangle$ &
$\langle\hat{F}_{\mu\nu}^{+}\hat{\chi}_{+}\rangle$ &
$[\hat{F}_{\mu\nu}^{+},\hat{\chi}_{+}]$ &
$\{\hat{F}_{\mu\nu}^{+},\hat{\chi}_{+}\}$\\
 \noalign{\smallskip}\hline
\end{tabular}
\end{table*}

\section{Formalism up to one-loop level}\label{SecAnaly}

We adopt the standard power counting scheme as in
Refs.~\cite{Ecker:1994gg,Meissner:1997ws}. The chiral order
$D_{\chi}$ of a Feynman diagram is
\begin{equation}
D_{\chi}=2L+1+\sum_d (d-2)N^M_d+\sum_d (d-1)N^{MB}_d, \label{pwct}
\end{equation}
with $L$ the number of loops and $N^M_d$, $N^{MB}_d$ the number of
the $d$ dimension vertices from the meson and meson-baryon
Lagrangians, respectively. The chiral order of the magnetic moment
$\mu_T$ is $(D_{\chi}-1)$.

\begin{figure}
\centering
\includegraphics[width=0.99\hsize]{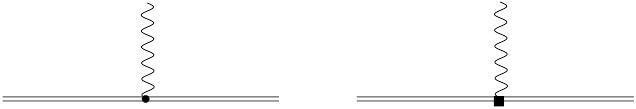}
\caption{The $\mathcal{O}(p^{2})$ and $\mathcal{O}(p^{4})$ tree
level diagrams where the spin-$3\over 2$ doubly charmed baryon is denoted by the
double solid line. The solid dot and black square represent the second- and
fourth-order couplings respectively.} \label{fig:tree}
\end{figure}

The leading order ($\mathcal{O}(p^1)$) magnetic moments come from
the tree diagram in Fig.~\ref{fig:tree} with the $\mathcal{O}(p^2)$
vertex. The magnetic moment is
\begin{equation}
\mu^{(1)}=2\alpha\frac{M_{N}}{M_{T}}\mu_{N}\label{mu1}
\end{equation}
The $\mu_N$ is the nucleon magneton. $\alpha$ are the Clebsch-Gordan
coefficients, which are collected in Table ~\ref{Table:cgloop}.
Up to the leading order, there are two unknown LECs, $b_1^{tt}$ and
$b_2^{tt}$.

\begin{figure*}[tbh]
\centering
\includegraphics[width=0.9\hsize]{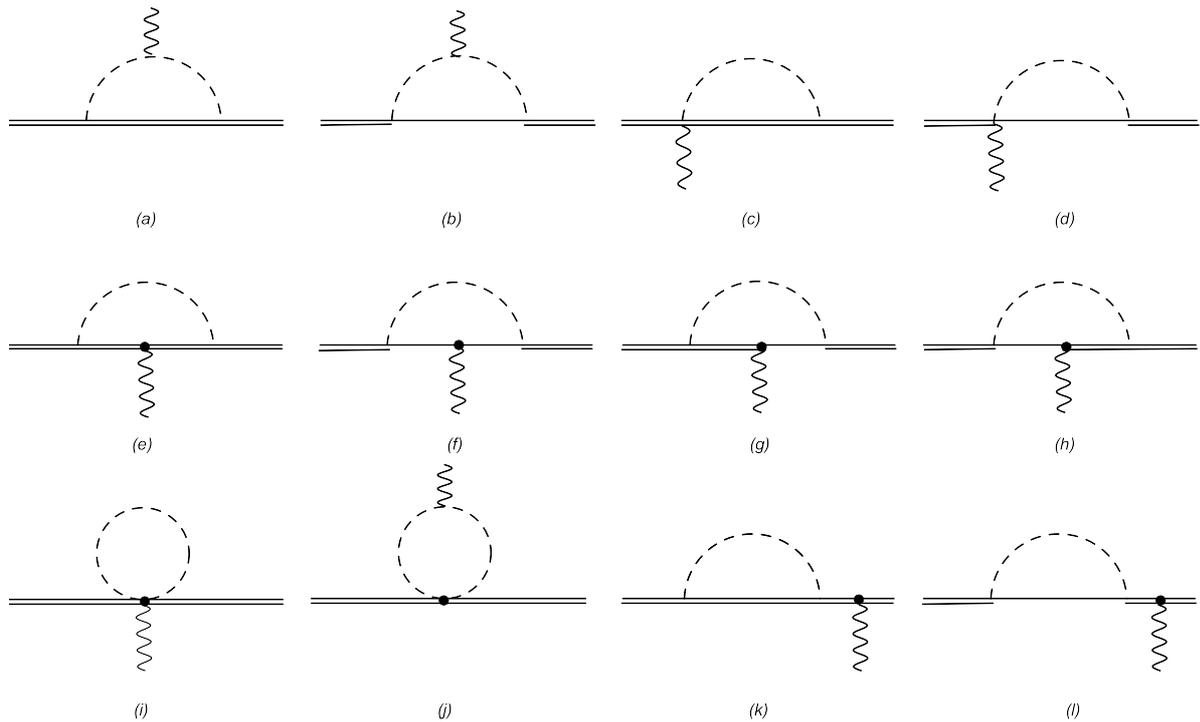}
\caption{The one-loop diagrams where the spin-$3\over 2$ (spin-$1\over 2$) doubly charmed baryon is
denoted by the double (single) solid line. The dashed and wiggly lines represent the
pseudoscalar meson and photon respectively. The solid dotes
represent $\mathcal{O}(p^2)$ vertices, while the other vertices are at the leading order.}\label{fig:allloop}
\end{figure*}

There are four diagrams (a)-(d) which contribute to the
next-to-leading order magnetic moments of the spin-$3 \over 2$ doubly charmed baryons, as shown in Fig.~\ref{fig:allloop}. All the
vertices in (a)-(d) come from Eq.~(\ref{Lag:meson1}) and
Eqs.~(\ref{Lag:op1rdtt})(\ref{Lag:op1rdbt}). The diagrams (c) and
(d) vanish in the heavy baryon mass limit. In particular, the
amplitudes of the diagrams (c) and (d) are denoted as
$\mathcal{M}_c$ and $\mathcal{M}_d$. We have
\begin{eqnarray}
 \mathcal{M}_c & \propto & \int\frac{d^{d}l}{(2\pi)^{d}}\frac{i}{l^{2}-m^{2}_\phi+i\epsilon}\frac{(S\cdot l)}{f_{0}}
 \frac{-iP_{\rho\sigma}^{3/2}}{v\cdot l+i\epsilon}S_\mu\protect\\
 \nonumber
 & \propto & S\cdot v=0,\\
\mathcal{M}_d & \propto &
\int\frac{d^{d}l}{(2\pi)^{d}}\frac{i}{l^{2}-m^{2}_\phi+i\epsilon}\frac{
l_\sigma}{f_{0}}
 \frac{i}{v\cdot l- \omega +i\epsilon}g_{\mu\rho}\protect\\  \nonumber
 & \propto & g_{\mu\rho}v_{\sigma},
 \end{eqnarray}
where $P_{\rho\sigma}^{3/2}$ is the non-relativistic spin-$\frac32$
projector. $\mathcal{M}_d$ vanishes since $v_{\sigma}u^{\sigma}=0$.
In other words, diagrams (c) and (d) do not contribute to the
magnetic moments in the leading order of the heavy baryon expansion.
The $\mathcal{O}(p^2)$ magnetic moment is
\begin{equation}
\mu^{(2)}=\sum_{\phi=\pi,K}({\cal A}^{\phi}+{\cal
B}^{\phi})\beta_{a}^{\phi},
\end{equation}
\begin{equation}
{\cal
A}^{\phi}=-\frac{H^{2}M_{N}m_{\phi}}{96f_{\phi}^{2}}\mu_{N},
\end{equation}
\begin{equation}
\begin{split}
{\cal
B}^{\phi}=-\frac{C^{2}M_{N}\mu_{N}}{64\pi^{2}f_{\phi}^{2}}[&2\sqrt{m_{\phi}^{2}-\delta^{2}}\arccos\left(\frac{\delta}{m_{\phi}}\right)\\
&-\delta\left(\ln\frac{m_{\phi}^{2}}{\lambda^{2}}-1\right)]
\end{split}
\end{equation}
where $\beta^\phi_a$ are the Clebsch-Gordan coefficients, which are
collected in Fig.~\ref{Table:cgloop}. Up to $\mathcal{O}(p^2)$,
there are five unknown LECs, $b_{1,2}^{tt}$, $C$, $H$ and
$\tilde{g}_A$. $C$, $H$ and $\tilde{g}_A$ can be estimated with the
quark model.

There are eight loop diagrams (e)-(l) in Fig.~\ref{fig:allloop},
which contribute at $\mathcal{O}(p^3)$. For the (e)-(h) diagrams,
the photon-baryon vertices are from the $\mathcal{O}(p^2)$
interaction terms in Eqs.~(\ref{Lag:op2rdtt}-\ref{Lag:op2rdbb}),
while the meson-baryon vertices are from Eqs.~(\ref{Lag:op1rdtt})(\ref{Lag:op1rdbt}). The vertex in diagram
(i) is from Eq.~(\ref{Lag:op2rdtt}). The meson-baryon vertex in
diagram (j) is from the Lagrangian in Eq.~(\ref{Lag:op2g}) while the
meson-photon vertex is from the Lagrangian in
Eq.~(\ref{Lag:meson1}). The loops (k) and (l) represent the wave
function renormalization. The photon-baryon vertices are from
Lagrangians in Eq.~(\ref{Lag:op2rdtt}), while the meson-baryon
vertices are from the interaction in
Eqs.~(\ref{Lag:op1rdtt})(\ref{Lag:op1rdbt}). The contributions to
the magnetic moments from the eight loop diagrams read:
\begin{equation}
\begin{split}
\mu^{(3)}_{\text {loop}}=&\sum_{\phi=\eta,\pi,K}({\cal
E}^{\phi}\gamma_{e}^{\phi}+{\cal F}^{\phi}\gamma_{f}^{\phi}+{\cal
G}^{\phi}\gamma_{g}^{\phi}+{\cal H}^{\phi}\gamma_{h}^{\phi})\\
&+\sum_{\phi=\pi,K}({\cal I}^{\phi}\delta^{\phi}+{\cal
J}^{\phi}\eta^{\phi}) +\sum_{\phi=\pi,\eta,K}\left({\cal
K}^{\phi}+{\cal L}^{\phi}\right)\xi^{\phi}\mu^{(1)}
\end{split}
\end{equation}
where
\begin{widetext}
\begin{eqnarray}
&&{\cal E}^{\phi}=\frac{H^{2}m_{\phi}^{2}}{864\pi^{2}f_{\phi}^{2}}\frac{M_{N}}{M_{T}}\mu_{N}\left(33\ln\frac{m_{\phi}^{2}}{\lambda^{2}}+70\right)\\
&&{\cal F}^{\phi}=\frac{C^{2}}{64\pi^{2}f_{\phi}^{2}}\frac{M_{N}}{M_{T}}\mu_{N}\left[2\delta^{2}+\left(m_{\phi}^{2}-2\delta^{2}\right)\ln\frac{m_{\phi}^{2}}{\lambda^{2}}+4\delta\sqrt{m_{\phi}^{2}-\delta^{2}}\arccos\left(\frac{\delta}{m_{\phi}}\right)\right]\\
&&{\cal G}^{\phi}={\cal H}^{\phi}=\frac{CH}{864\pi^{2}f_{\phi}^{2}\delta}\frac{M_{N}}{M_{T}}\mu_{N}\left[-2\left(\delta^{3}+3\pi m_{\phi}^{3}\right)+\left(6\delta^{3}-9\delta m_{\phi}^{2}\right)\ln\left(\frac{m_{\phi}^{2}}{\lambda^{2}}\right)+12\left(m_{\phi}^{2}-\delta^{2}\right)^{3/2}\arccos\left(\frac{\delta}{m_{\phi}}\right)\right]\\
&&{\cal I}^{\phi}={\cal J}^{\phi}=\frac{m_{\phi}^{2}}{64\pi^{2}f_{\phi}^{2}}\frac{M_{N}}{M_{T}}\text{ln}\frac{m_{\phi}^{2}}{\lambda^{2}}\mu_{N}\\
&&{\cal K}^{\phi}=-\frac{H^{2}m_{\phi}^{2}}{576\pi^{2}f_{\phi}^{2}}\left(15\ln\frac{m_{\phi}^{2}}{\lambda^{2}}+26\right)\\
&&{\cal
L}^{\phi}=\frac{-C^{2}}{64\pi^{2}f_{\phi}^{2}}\left[2\delta^{2}+\left(m_{\phi}^{2}-2\delta^{2}\right)\ln\frac{m_{\phi}^{2}}{\lambda^{2}}+4\delta\sqrt{m_{\phi}^{2}-\delta^{2}}\arccos\left(\frac{\delta}{m_{\phi}}\right)\right]
\end{eqnarray}
\end{widetext}
The $\gamma^{\phi}_{e-h}$,$\delta^{\phi}$, $\eta^{\phi}$ and
$\xi^{\phi}$ are loop coefficients, which are listed in
Table~\ref{Table:cgloop}.

Apart from the loop diagrams, there is a tree diagram which
contributes to the $\mathcal{O}(p^3)$ magnetic moments. The
Lagrangian is given in Eq.~\ref{op4r}. The contribution reads
\begin{equation}
\mu_{\text{\text{tree}}}^{(3)}=2\phi\frac{M_{N}}{M_{T}}\mu_{N}
\end{equation}
where the coefficients $\phi$ are given in Table~\ref{Table:cgtree}.
Up to $\mathcal{O}(p^3)$ magnetic moments, there are thirteen LECs,
$\tilde{g}_A$, $C$, $H$, $b_{1,2}^{tt,bb,bt}$, $g_2^{tt}$, and
$a_{1,2,3}$.

\section{Numberical results and discussions}\label{SecNO}

Since the spin-$3 \over 2$ doubly charmed baryons have not been
observed in the experiments, we do not have any experimental inputs
to fit the LECs. However, we can employ quark model to determine
some of the LECs. The numerical results are given in Table
\ref{Table:NO}.

There are two unknown LECs $b_{1,2}^{tt}$ up to the leading order
($\mathcal{O}(p^1)$) magnetic moments. Unlike the light baryons, the
charge matrix $Q_{B}$ of the doubly charmed baryons is not
traceless. The heavy quarks contribute to the trace part of the
charge matrix. Thus, in the second column of Table~\ref{Table:NO},
the $b_{1}^{tt}$ terms come from the light quark contribution. The $b_2^{tt}$ terms are the same for
the three doubly charmed baryons and arise solely from the two charm
quarks.

At the quark level, the flavor and spin wave function of the
$\Xi_{cc}^{*++}$ reads:
\begin{eqnarray}
|\Xi_{cc}^{*++};S_3={3\over 2}\rangle =|c\uparrow c\uparrow
u\uparrow\rangle, \label{xiwavefunc}
\end{eqnarray}
where the arrow denote the third component of the spin. The
magnetic moments of the baryons in the quark model are the matrix
elements of the following operator $\mu$ sandwiched between the wave functions,
\begin{equation}
\mu=\sum_i\mu_i\sigma_3^i, \label{magmomen}
\end{equation}
where $\mu_i$ is the magnetic moment of the quark.
\begin{equation}
\mu_i={e_i\over 2m_i},\quad i=u,d,s,c,b.
\end{equation}
We adopt the constituent quark masses from Ref.\cite{Lichtenberg:1976fi} as the set A
with $m_u=m_d=336$ MeV, $m_s=540$ MeV, $m_c=1660$ MeV and $m_b=4700$
MeV. We adopt the constituent quark masses from Ref.~\cite{Karliner:2014gca} as the set
B with $m_u=m_d=363$ MeV, $m_s=538$ MeV, $m_c=1711$ MeV and
$m_b=5044$ MeV. The magnetic moments from the naive quark model
estimation with set A and B are given in Table~\ref{quark model}. The two sets lead
to the similar magnetic moments. We choose the results of set A in
the following calculation.
\begin{table}
\caption{The numberical results of spin-$3\over 2$ doubly heavy baryon magnetic moments from naive quark model estimation. Set A and set B are two constituent quark masses sets from Refs.~\cite{Lichtenberg:1976fi,Karliner:2014gca}, respectively.}\label{quark model}
\centering
\begin{tabular}{l|c|cc}
\hline\noalign{\smallskip}
Baryons & Magnetic moments & Set A & Set B\\
 \noalign{\smallskip}\hline\noalign{\smallskip} 
$\Xi_{cc}^{*++}$ & $2\mu_{c}+\mu_{u}$ & 2.61 & 2.45\\
$\Xi_{cc}^{*+}$ & $2\mu_{c}+\mu_{d}$ & -0.18 & -0.13\\
$\Omega_{cc}^{*+}$ & $2\mu_{c}+\mu_{s}$ & 0.17 & 0.15\\
$\Xi_{bb}^{*0}$ & $2\mu_{b}+\mu_{u}$ & 1.73 & 1.60\\
$\Xi_{bb}^{*-}$ & $2\mu_{b}+\mu_{d}$ & -1.06 & -0.99\\
$\Omega_{bb}^{*-}$ & $2\mu_{b}+\mu_{s}$ & -0.71 & -0.71\\
$\Xi_{bc}^{*+}$ & $\mu_{b}+\mu_{c}+\mu_{u}$ & 2.17 & 2.03\\
$\Xi_{bc}^{*0}$ & $\mu_{b}+\mu_{c}+\mu_{d}$ & -0.62 & -0.56\\
$\Omega_{bc}^{*0}$ & $\mu_{b}+\mu_{c}+\mu_{s}$ & -0.27 & -0.28\\ \noalign{\smallskip}\hline
\end{tabular}
\end{table}


The $\mathcal{O}(p^1)$ magnetic moments
of the spin-$3\over 2$ doubly charmed baryons are given in the
second column in Table~\ref{Table:NO}. The numerical results from the quark model are given in the third column. In the quark model, the light
quark parts contribute to the $b_{1}^{tt}$ terms, which are proportional to the light quark charge. The heavy quark parts
contribute to the $b_{2}^{tt}$ terms, which are the same for the
three doubly charmed baryons.

Up to $\mathcal{O}(p^{2})$, we must take the loop corrections into
consideration. At this order, there exist three new LECs
$\tilde{g}_A$, $C$ and $H$, which are estimated in the quark
model~\cite{Li:2017cfz,Li:2017pxa}. The numerical results are given
in the third column of Table~\ref{Table:NO}.

In Table~\ref{Table:NO}, the magnetic moments of the
$\Xi^{*++}_{cc}$ and $\Xi^{*+}_{cc}$ are dominated by the leading
order term while the magnetic moment of $\Omega_{cc}^{*+}$ is
dominated by the chiral corrections. At the leading order, since
three quarks in $\Xi_{cc}^{*++}$ all have the positive charge, their
contributions to the magnetic moments are constructive. For the
$\Xi_{cc}^{*+}$ and $\Omega_{cc}^{*+}$, the contribution of the
heavy quarks and light quark cancel out to a large extent, which
leads to small magnetic moments at the leading order.

At the next-to-leading order, both $\pi^+$ and $K^+$ mesons
contribute to the chiral correction of $\mu_{\Xi^{*++}_{cc}}$, while
only $\pi^+$ ($K^+$) contributes to $\mu_{\Xi^{*+}_{cc}}$
($\mu_{\Omega^{*+}_{cc}}$). The chiral corrections in the loops (a)
and (b) are proportional to the pseudoscalar meson mass, i.e., $\sim{
m_\phi \over M_T}$. Therefore, the chiral corrections for
$\mu_{\Xi^{*++}_{cc}}$ and $\mu_{\Omega_{cc}^{*+}}$ are much larger than
that for $\mu_{\Xi_{cc}^{*+}}$. It is interesting to note that the
chiral correction from the $K^+$ loop for the
$\mu_{\Omega_{cc}^{*+}}$ is much lager than the leading order
contribution. Such a unique feature might be exposed by future
lattice QCD simulation.

Up to $\mathcal{O}(p^{3})$, eight new LECs, $b_{1,2}^{bb,bt}$,
$g_2^{tt}$ and $a_{1,2,3}$ are introduced. Since it is impossible to
fix all these LECs due to lack of experimental data, we do not
present the numerical results up to this order.

\begin{table*}
\caption{The spin-$3\over 2$ doubly charmed baryon magnetic moments
when the chiral expansion is truncated at $\mathcal{O}(p^{1})$ and
$\mathcal{O}(p^{2})$, respectively (in unit of $\mu_{N}$). The
magnetic moments in the third column are evaluated by quark model,
which are treated as the $\mathcal{O}(p^1)$ magnetic moments. We adopt $\delta=-100$ MeV and $\delta=-70$ MeV as two parameter sets.}
\label{Table:NO}
  \centering
\begin{tabular}{l|cc|ccc|ccc}
\hline\noalign{\smallskip}
Baryons & ${\cal O}(p^{1})$ & quark model & $\delta$/MeV & ${\cal O}(p^{2})$ I & Total I & $\delta$/MeV & ${\cal O}(p^{2})$ II & Total II\\
 \noalign{\smallskip}\hline\noalign{\smallskip}
$\Xi_{cc}^{*++}$ & $-\frac{4}{3}\frac{M_{N}}{M_T}b_{1}^{tt}-\frac{8}{3}\frac{M_{N}}{M_T}b_{2}^{tt}$ & $2\mu_{c}+\mu_{u}=2.61$ & -100 & -0.90 & 1.72 & -70 & -1.02 & 1.59\\
$\Xi_{cc}^{*+}$ & $\frac{2}{3}\frac{M_{N}}{M_T}b_{1}^{tt}-\frac{8}{3}\frac{M_{N}}{M_T}b_{2}^{tt}$ & $2\mu_{c}+\mu_{d}=-0.18$ & -100 & 0.09 & -0.09 & -70 & 0.19 & 0.02\\ 
$\Omega_{cc}^{*+}$ & $\frac{2}{3}\frac{M_{N}}{M_T}b_{1}^{tt}-\frac{8}{3}\frac{M_{N}}{M_T}b_{2}^{tt}$ & $2\mu_{c}+\mu_{s}=0.17$ & -100 & 0.81 & 0.99 & -70 & 0.82 & 1.00\\
 \noalign{\smallskip}\hline
\end{tabular}
\end{table*}

With the same formalism, we have also calculated the magnetic
moments of the doubly bottom baryons and charmed bottom baryons. Since the $b$ quark is much heavier than the $c$ quark, we adopt
the mass splitting $\delta=-40$ MeV and $\delta=-20$ MeV for the
doubly bottom baryons and $\delta=-60$ MeV and $\delta=-40$ MeV for
the charmed bottom baryons. The above mass difference in our work is
consistent with that in Refs.~\cite{Martynenko:2007je,Patel:2007gx,Karliner:2014gca,Shah:2016vmd}. The leading order magnetic
moments and the LECs in the next-to-leading order magnetic moments
are estimated by the quark model. We present the numerical results
of the doubly bottom baryon and charmed bottom baryons magnetic
moments to next-to-leading order in Table \ref{Table:mubbq}.

\begin{table*}
\caption{The spin-$3\over 2$ doubly bottom baryon (bbq) and charmed
bottom baryon (bcq) magnetic moments to the next-to-leading order
(in unit of $\mu_N$ ). We use the subscripts ${bc}$ to label the
systems with symmetric spin wave functions in the heavy quark
sector. We adopt $\delta=-40$ and $\delta=-20$ MeV for the doubly bottom baryons and $\delta=-60$ and $\delta=-20$ MeV for the charmed bottom baryons} \label{Table:mubbq}
\centering
\begin{tabular}{l|c|ccc|ccc}
\hline\noalign{\smallskip}
Baryons & ${\cal O}(p^{1})$ & $\delta$/MeV & ${\cal O}(p^{2})$ I & Total I & $\delta$/MeV & ${\cal O}(p^{2})$ II  & Total II\\
 \noalign{\smallskip}\hline\noalign{\smallskip}
$\Xi_{bb}^{*0}$ & $2\mu_{b}+\mu_{u}=1.73$ & -40 & -1.10 & 0.63 & -20 & -1.15 & 0.58\\
$\Xi_{bb}^{*-}$ & $2\mu_{b}+\mu_{d}=-1.06$ & -40 & 0.27 & -0.79 & -20 & 0.31 & -0.75\\
$\Omega_{bb}^{*-}$ & $2\mu_{b}+\mu_{s}=-0.71$ & -40 & 0.83 &0.12 & -20 & 0.83 & 0.12\\
 \noalign{\smallskip}\hline\noalign{\smallskip}
$\Xi_{bc}^{*+}$ & $\mu_{b}+\mu_{c}+\mu_{u}=2.17$ & -60 & -1.05 & 1.12 & -40 & -1.10 & 1.07\\
$\Xi_{bc}^{*0}$ & $\mu_{b}+\mu_{c}+\mu_{d}=-0.62$ & -60 & 0.22 & -0.40 & -40 & 0.27 & -0.35\\
$\Omega_{bc}^{*0}$ & $\mu_{b}+\mu_{c}+\mu_{s}=-0.27$ & -60 & 0.82 & 0.56 & -40 & 0.83 & 0.56\\
 \noalign{\smallskip}\hline
\end{tabular}
\end{table*}

\section{Conclusions}\label{SecConcld}

The doubly heavy baryons are particularly interesting because their
chiral dynamics is solely dominated by the single light quark. In
this work, we have employed the heavy baryon chiral perturbation
theory to calculate the magnetic moments of the spin-$3 \over 2$
doubly charmed baryons, which reveal the information of their inner
structure. Due to the large mass of the doubly heavy baryons, the
recoil corrections are expected to be very small. We have derived
our analytical expressions up to the next-to-next-to-leading order,
which may be useful to the possible chiral extrapolation of the
lattice simulations of the doubly charmed baryon electromagnetic
properties.

With the help of quark model, we have estimated the LECs and
presented the numerical results up to next-to-leading order:
$\mu_{\Xi^{*++}_{cc}}=1.72\mu_{N}$,
$\mu_{\Xi^{*+}_{cc}}=-0.09\mu_{N}$,
$\mu_{\Omega^{*+}_{cc}}=0.99\mu_{N}$ for $\delta=-100$ MeV and $\mu_{\Xi^{*++}_{cc}}=1.59\mu_{N}$,
$\mu_{\Xi^{*+}_{cc}}=0.02\mu_{N}$,
$\mu_{\Omega^{*+}_{cc}}=1.00\mu_{N}$ for $\delta=-70$ MeV. As by-products, we have also
calculated the magnetic moments of the spin-$3\over 2$ doubly bottom
baryons and charmed bottom baryons.

For comparison, we have listed the spin-$3\over 2$ doubly charmed
baryon magnetic moments from some other model calculations in
Table~\ref{Table:Comp} including quark
model~\cite{Lichtenberg:1976fi}, MIT bag
model~\cite{Bose:1980vy,Bernotas:2012nz}, Skymion
model~\cite{Oh:1991ws}, nonrelativistic quark model
(NRQM)~\cite{Albertus:2006ya}, hyper central quark model
(HCQM)~\cite{Patel:2007gx}, effective mass and screened charge
scheme~\cite{Dhir:2009ax} and chiral constituent quark model
($\chi$CQM)~\cite{Sharma:2010vv}. We define the relative changes $\rho$ of these models with respect
to the naive estimates as $\mu=(1+\rho)\mu_{\text{QM}}$, where
$\mu_{\text{QM}}$ is the magnetic moments estimated with the naive
quark model. For the $\Xi_{cc}^{*++}$ baryon, one notices that the
$|\rho|<0.1$ for most models, and $|\rho|<0.4$ for all models. Thus,
all these approaches lead to more or less similar numerical results
for the magnetic moments of $\Xi_{cc}^{*++}$. For the
$\Xi_{cc}^{*+}$ baryon, the $|\rho|<2.5$ for most models including
our work while the $|\rho|>4$ from the Skymion models. For the
$\Omega_{cc}^{*+}$ baryon, the $|\rho|< 2$ in all the models except
our work and one Skymion calculation. Thus, various models lead to
quite different predictions for the magnetic moments of the
$\Xi_{cc}^{*+}$ and $\Omega_{cc}^{*+}$, which may be used to
distinguish these models.

In the numerical analysis, we have truncated the chiral expansions
at $\mathcal{O}(p^{2})$ and omitted all the $\mathcal{O}(p^{3})$
higher order chiral corrections because of too many unknown LECs at
this oder. When more experimental measurements become available in
the future, the numerical analysis in the present work can be
further improved. In principle, all the low energy constants shall
be extracted through fitting to the experimental data instead of
using the estimation from the quark model. The $\mathcal{O}(p^{3})$
chiral corrections may turn out to be non-negligible and should be
included.

There is good hope that the spin-$3\over 2$ doubly charmed baryons
will be observed through its radiative or weak decays in the coming
future. We hope our numerical calculation may be useful for future
experimental measurements of their magnetic moments. There are
several LECs in our analytical results to be determined by the
future progresses in the experiment and theory, which will help
check the chiral expansion convergence of the three doubly charmed
baryons.

\begin{table}
\caption{Comparison of the spin-$3 \over 2$ doubly charmed baryons
magnetic moments in the literature including quark
model~\cite{Lichtenberg:1976fi}, MIT bag
model~\cite{Bose:1980vy,Bernotas:2012nz}, Skymion
model~\cite{Oh:1991ws}, nonrelativistic quark model
(NRQM)~\cite{Albertus:2006ya}, hyper central model
(HCQM)~\cite{Patel:2007gx}, effective mass and screened charge
scheme~\cite{Dhir:2009ax}. We adopt $\delta=-100$ MeV and $\delta=-70$ MeV as I and II parameter sets respectively in this work.} \label{Table:Comp}
\centering
\begin{tabular}{l|ccc}
\hline\noalign{\smallskip}
 & $\Xi_{cc}^{*++}$  & $\Xi_{cc}^{*+}$ &  $\Omega_{cc}^{*+}$ \\
 \noalign{\smallskip}\hline\noalign{\smallskip}
Quark model~\cite{Lichtenberg:1976fi}  & 2.60 &  -0.19 &  0.17 \\
Bag model l~\cite{Bose:1980vy} & 2.54  & 0.20  & 0.39\\
Bag model 2~\cite{Bernotas:2012nz} & 2.00 & 0.16 &  0.33 \\
Skymion 1~\cite{Oh:1991ws}
& 3.16 &  -0.98 &  -0.20 \\ 
Skymion 2~\cite{Oh:1991ws} & 3.18 & -1.17 & 0.03\\
NRQM~\cite{Albertus:2006ya} & 2.67 & -0.31  & 0.14 \\
HCQM~\cite{Patel:2007gx} & 2.75 & -0.17  & 0.12\\ 
Effective mass~\cite{Dhir:2009ax} & 2.41 & -0.11 &  0.16 \\
Screened charge~\cite{Dhir:2009ax}
& 2.52 & 0.04  & 0.21 \\
$\chi\text{CQM}$~\cite{Sharma:2010vv} & 2.66& -0.47  & 0.14 \\
This work I & 1.72  & -0.09  & 0.99 \\
This work II & 1.59  & 0.02  & 1.00 \\
\noalign{\smallskip}\hline
\end{tabular}
\end{table}

\begin{acknowledgements}
L. Meng is very grateful to X. L. Chen and W. Z. Deng for very
helpful discussions. This project is supported by the National
Natural Science Foundation of China under Grants 11575008,
11621131001 and 973 program. This work is also supported by the
Fundamental Research Funds for the Central Universities of Lanzhou
University under Grants 223000--862637.
\end{acknowledgements}

\begin{appendix}

\section{COEFFICIENTS OF THE LOOP CORRECTIONS} \label{AppCG}

In this appendix, we collect the explicit formulae for the chiral
expansion of the doubly charmed baryon magnetic moments in Tables
\ref{Table:cgtree} and \ref{Table:cgloop}.

\begin{table*}
  \centering
  \caption{The coefficients of the loop corrections to the
spin-${3\over 2}$ doubly charmed baryon magnetic moments from
Fig.~\ref{fig:allloop}. For diagrams (e)-(h), the LECs $b_{1,2}$ in
the Table represent $b_{1,2}^{tt}$, $b_{1,2}^{bb}$, $b_{1,2}^{bt}$
and $b_{1,2}^{bt}$ in order. } \label{Table:cgloop}
\begin{tabular}{lcc|ccc|cc|cc|ccc}
\hline\noalign{\smallskip}
 & $\beta_{a}^{\pi}$ & $\beta_{a}^{K}$ & $\gamma_{e-h}^{\pi}$ & $\gamma_{e-h}^{K}$ & $\gamma_{e-h}^{\eta}$ & $\delta^{\pi}$ & $\delta^{K}$ & $\eta^{\pi}$ & $\eta^{K}$ & $\xi^{\pi}$ & $\xi^{K}$ & $\xi^{\eta}$\\
 \noalign{\smallskip}\hline\noalign{\smallskip}
$\Xi_{cc}^{*++}$ & $4$ & $4$ & $4b_{2}$ &
$\frac{2}{3}\left(-b_{1}+4b_{2}\right)$ &
$\frac{2}{9}\left(b_{1}+2b_{2}\right)$ & $-4b_{1}^{tt}$ &
$-4b_{1}^{tt}$ & $-8g_{2}^{tt}$ & $-8g_{2}^{tt}$ & $3$ & $2$ &
$\frac{1}{3}$\\
 $\Xi_{cc}^{*+}$ & $-4$ &  &
$b_{1}+4b_{2}$ & $\frac{2}{3}\left(-b_{1}+4b_{2}\right)$ &
$\frac{1}{9}\left(-b_{1}+4b_{2}\right)$ & $4b_{1}^{tt}$ &  &
$8g_{2}^{tt}$ &  & $3$ & $2$ & $\frac{1}{3}$\\
$\Omega_{cc}^{*+}$ &  & $-4$ &  &
$\frac{2}{3}\left(b_{1}+8b_{2}\right)$ &
$\frac{4}{9}\left(-b_{1}+4b_{2}\right)$ &  & $4b_{1}^{tt}$ &  &
$8g_{2}^{tt}$ &  & $4$ & $\frac{4}{3}$\\
 \noalign{\smallskip}\hline
\end{tabular}
\end{table*}

\begin{table}
\caption{The coefficients of the tree diagrams in
Fig.\ref{fig:tree}, which contribute to the $\mathcal{O}(p^1)$ and
$\mathcal{O}(p^3)$ magnetic moments. } \label{Table:cgtree}
\centering
\begin{tabular}{lc|c}
\hline\noalign{\smallskip}
 & $\alpha$ & $\phi$\\
 \noalign{\smallskip}\hline\noalign{\smallskip}
$\Xi_{cc}^{*++}$ & $\frac{2}{3}b_{1}^{tt}+\frac{4}{3}b_{2}^{tt}$ &
$-\frac{4}{9}a_{1}-\frac{1}{9}a_{2}-\frac{4}{9}a_{3}$\\
$\Xi_{cc}^{*+}$ & $-\frac{1}{3}b_{1}^{tt}+\frac{4}{3}b_{2}^{tt}$ &
$-\frac{4}{9}a_{1}-\frac{1}{9}a_{2}+\frac{2}{9}a_{3}$\\
$\Omega_{cc}^{*+}$ & $-\frac{1}{3}b_{1}^{tt}+\frac{4}{3}b_{2}^{tt}$
&
$\frac{8}{9}a_{1}-\frac{1}{9}a_{2}-\frac{4}{9}a_{3}$\\
 \noalign{\smallskip}\hline
\end{tabular}
\end{table}
\end{appendix}

\end{document}